\documentclass[10pt,a4paper,twocolumn]{article}

\usepackage{ifthen} 
\usepackage{overpic}
\newboolean{pdflatex}
\setboolean{pdflatex}{true} 

\newboolean{articletitles}
\setboolean{articletitles}{true} 

\newboolean{uprightparticles}
\setboolean{uprightparticles}{false} 

\newboolean{inbibliography}
\setboolean{inbibliography}{false} 

\usepackage{microtype}
\usepackage{lineno}  
\usepackage{xspace} 

\usepackage{graphicx}  
\usepackage{color}
\usepackage{colortbl}
\graphicspath{{./figs/}} 

\usepackage{amsmath} 
\usepackage{amssymb}
\usepackage{amsfonts}
\usepackage{upgreek} 

\newcommand*\patchAmsMathEnvironmentForLineno[1]{%
\expandafter\let\csname old#1\expandafter\endcsname\csname #1\endcsname
\expandafter\let\csname oldend#1\expandafter\endcsname\csname
end#1\endcsname
 \renewenvironment{#1}%
   {\linenomath\csname old#1\endcsname}%
   {\csname oldend#1\endcsname\endlinenomath}%
}
\newcommand*\patchBothAmsMathEnvironmentsForLineno[1]{%
  \patchAmsMathEnvironmentForLineno{#1}%
  \patchAmsMathEnvironmentForLineno{#1*}%
}
\AtBeginDocument{%
\patchBothAmsMathEnvironmentsForLineno{equation}%
\patchBothAmsMathEnvironmentsForLineno{align}%
\patchBothAmsMathEnvironmentsForLineno{flalign}%
\patchBothAmsMathEnvironmentsForLineno{alignat}%
\patchBothAmsMathEnvironmentsForLineno{gather}%
\patchBothAmsMathEnvironmentsForLineno{multline}%
}

\usepackage{hyperref}    
\usepackage[all]{hypcap} 

\usepackage{cite} 
\usepackage{mciteplus}
\usepackage{multirow}

\hypersetup{%
  colorlinks=false, linktocpage=false, pdfborder={0 0 0}, pdfstartpage=1, 
 pdfstartview=FitV,%
}

\graphicspath{{./additional_material/}}

\newcommand{\BRof}[1]{\ensuremath{{\cal B}(#1)}\xspace}
\newcommand{\Bsmumu}{\ensuremath{B^0_s \to\mu^+\mu^-}\xspace}
\newcommand{\Bdmumu}{\ensuremath{B^0\to\mu^+\mu^-}\xspace}

\newcommand{\Bsmm}{\ensuremath{B^0_s\to\mu^+\mu^-}\xspace}

\def\bdkpi{\ensuremath{B^0 \to K^+ \pi^-}\xspace}
\newcommand{\Bdpipi}{\ensuremath{\Bd\to\pi^+\pi^-}\xspace}
\newcommand{\bpimumu}{\ensuremath{B^{0(+)} \to \pi^{0(+)} \mu^+ \mu^-}\xspace}
\newcommand{\BdPiMuNu}{\ensuremath{\ensuremath{B^0}\to \pi^- \mu^+ \nu_\mu}\xspace}
\def\BsKMuNu{\ensuremath{B^0_s \to K^- \mu^+ \nu_{\mu}}\xspace}
\mathchardef\PLambda="7103                 
\def\L {\ensuremath{\PLambda}\xspace}
\def\Lb{\ensuremath{\L_b^0}\xspace}

\def\Lbpmunu{\ensuremath{\Lb \to p \mu^- \bar \nu_{\mu}}\xspace}

\def\B       {\ensuremath{B}\xspace}
\def\Bbar    {\kern 0.18em\overline{\kern -0.18em \B}{}\xspace}
\def\Bs      {\ensuremath{B^0_s}\xspace}

\def\B       {\ensuremath{B}\xspace}
\def\Bd      {\ensuremath{B^0}\xspace}
\def\Bs      {\ensuremath{B^0_s}\xspace}
\newcommand{\mBd}{\ensuremath{m_{\Bd}}\xspace}
\newcommand{\mBs}{\ensuremath{m_{\Bs}}\xspace}

\def\lhcb {LHCb\xspace}
\newcommand{\Bhh}{\ensuremath{B^0_{(s)}\to h^+{h}^{\prime -}}\xspace}

\newcommand{\Bmumu}{\ensuremath{B^0_{(s)}\to \mu^+\mu^-}\xspace}
\newcommand{\BuJpsiK}{\ensuremath{B^+\to J/\psi K^+}\xspace}
\def\bujpsik{\BuJpsiK}
\newcommand{\BsJpsiPhi}{\ensuremath{B^0_s\to J/\psi \phi}\xspace}
\newcommand{\BdKpi}{\ensuremath{B^0\to K^+\pi^-}\xspace}
\newcommand{\BsKK}{\ensuremath{B^0_s\to K^+K^-}\xspace}
\newcommand{\bbdim}{\ensuremath{b\bar{b}\to \mu^+ \mu^- X}\xspace}
\newcommand{\CLsb}{\ensuremath{\textrm{CL}_{\textrm{s+b}}}\xspace}
\newcommand{\CLs}{\ensuremath{\textrm{CL}_{\textrm{s}}}\xspace}
\newcommand{\CLb}{\ensuremath{\textrm{CL}_{\textrm{b}}}\xspace}

\newcommand{\gevc}{\ensuremath{{\mathrm{\,Ge\kern -0.1em V\!/}c}}\xspace}
\newcommand{\mevc}{\ensuremath{{\mathrm{\,Me\kern -0.1em V\!/}c}}\xspace}
\newcommand{\gevcc}{\ensuremath{{\mathrm{\,Ge\kern -0.1em V\!/}c^2}}\xspace}
\newcommand{\gevgevcccc}{\ensuremath{{\mathrm{\,Ge\kern -0.1em V^2\!/}c^4}}\xspace}
\newcommand{\mevcc}{\ensuremath{{\mathrm{\,Me\kern -0.1em V\!/}c^2}}\xspace}

\def\Y#1S{\ensuremath{\Upsilon{(#1S)}}\xspace}

\newcommand\Tstrut{\rule{0pt}{2.6ex}}

\newcommand\Bstrut{\rule[-1.2ex]{0pt}{0pt}}

\def\gauss      {\mbox{\textsc{Gauss}}\xspace}
\def\pT         {\ensuremath{p_{\rm{T}}}\xspace}

\def\invfb      {\ensuremath{\mbox{\,fb}^{-1}}\xspace}

\newcommand{\tev}{\ensuremath{\mathrm{\,Te\kern -0.1em V}}\xspace}
\newcommand{\CL}{CL\xspace}
\def\BDT{BDT\xspace}

\newcommand{\tabcaption}[1]{
\vspace{-\abovecaptionskip}%
\caption{#1}
\vspace{0.5\abovecaptionskip}
}

\def\Bdexpsmn{\ensuremath{4.5 \times 10^{-10}}\xspace}
\def\Bdexpsmnf{\ensuremath{5.4 \times 10^{-10}}\xspace}

\def\Bdexpbkgn{\ensuremath{3.5 \times 10^{-10}}\xspace}
\def\Bdexpbkgnf{\ensuremath{4.4 \times 10^{-10}}\xspace}

\def\Bdobslimitnf{\ensuremath{7.4\times 10^{-10}}\xspace} 
\def\Bdobslimitn{\ensuremath{6.3\times 10^{-10}}\xspace}  
\def\Bdsigma{\ensuremath{2.0\,\sigma}\xspace}

\def\Bssign{\ensuremath{4.0}\xspace}
\def\Bssigma{\ensuremath{4.0\,\sigma}\xspace}
\def\Bsexpsigma{\ensuremath{5.0\,\sigma}\xspace}

\def\Bsbr{\ensuremath{(2.9^{\,+1.1}_{\,-1.0}
({\rm stat})^{\,+0.3}_{\,-0.1} ({\rm syst})) \times 10^{-9}}\xspace}
\def\Bsbrshort{\ensuremath{(2.9^{\,+1.1}_{\,-1.0})\times 10^{-9}}\xspace}

\def\Bdbr{\ensuremath{(3.7^{\,+2.4}_{\,-2.1}
({\rm stat})^{\,+0.6}_{\,-0.4} ({\rm syst})) \times 10^{-10}}\xspace}

\begin{document}


\onecolumn
\begin{titlepage}
\pagenumbering{roman}

\vspace*{-1.5cm}
\centerline{\large EUROPEAN ORGANIZATION FOR NUCLEAR RESEARCH (CERN)}
\vspace*{1.5cm}
\hspace*{-0.5cm}
\begin{tabular*}{\linewidth}{lc@{\extracolsep{\fill}}r}
\vspace*{-2.7cm}\mbox{\!\!\!\includegraphics[width=.14\textwidth]
{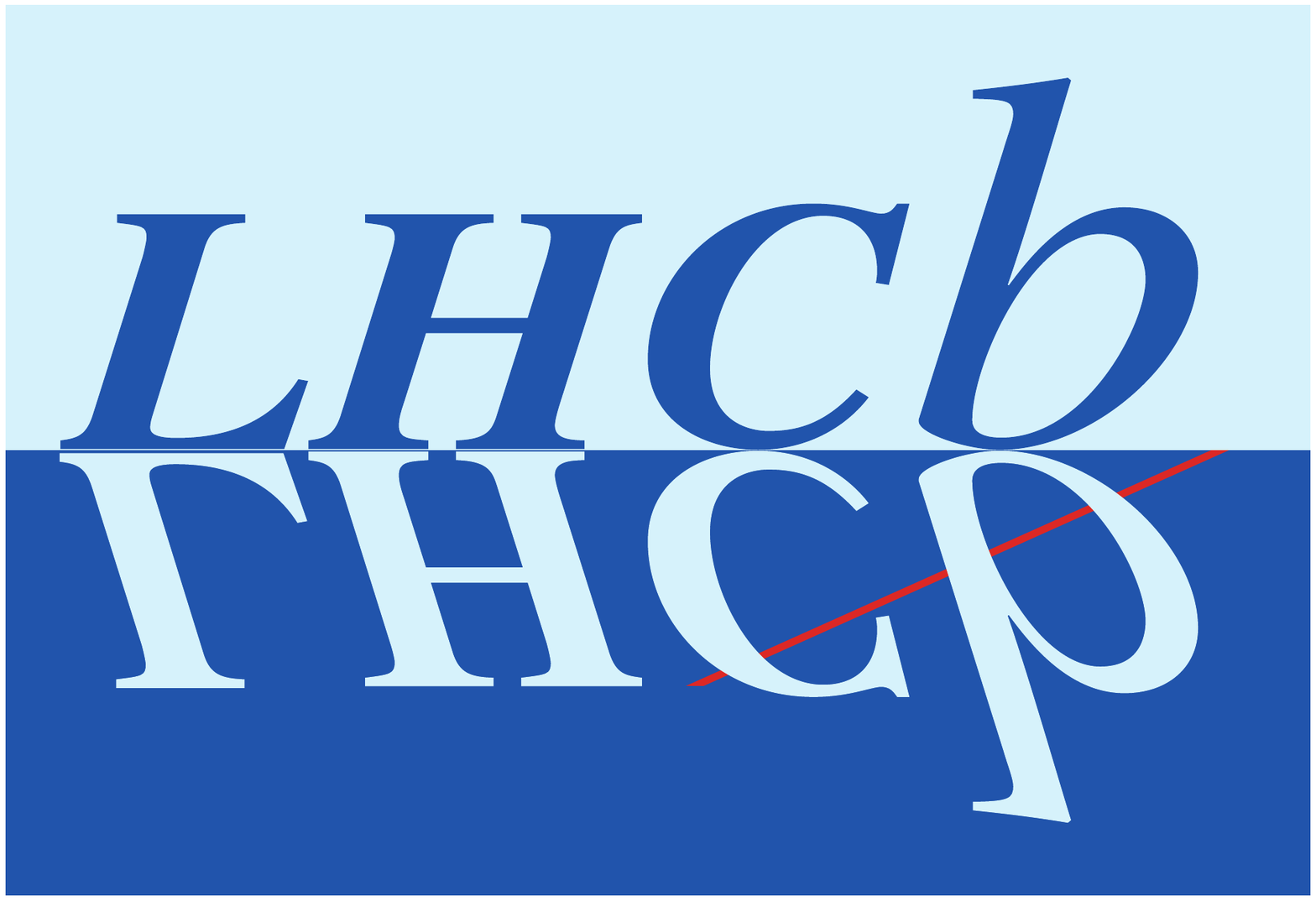}} & &
\\
 & & CERN-PH-EP-2013-128 \\  
 & & LHCb-PAPER-2013-046 \\  
 & & July 18, 2013 \\ 
 & & \\
\end{tabular*}

\vspace*{4.0cm}

{\bf\boldmath\huge
\begin{center}
Measurement of the \Bsmumu branching fraction and search for \Bdmumu 
decays\\ at the LHCb experiment 
\end{center}
}

\vspace*{2.0cm}

\begin{center}
The LHCb collaboration\footnote{Authors are listed on the following pages.}
\end{center}

\vspace{\fill}

\begin{abstract}
\noindent A search for the rare decays \Bsmumu and \Bdmumu is performed at the 
LHCb experiment. The data analysed 
correspond to an integrated luminosity of 1\invfb of $pp$ collisions at a 
centre-of-mass energy of $7$\tev and 2\invfb at $8$\tev.  
An excess of \Bsmumu signal candidates with respect to the background
expectation is seen with a significance of \Bssign standard 
deviations.
A time-integrated branching fraction of \BRof \Bsmumu = \Bsbrshort is obtained 
and an upper limit of \BRof \Bdmumu $< \Bdobslimitnf$ at 95\,\% confidence level
is set. 
These results are consistent with the Standard Model expectations.
\end{abstract}
\vspace*{\fill}

\begin{center}
  Published in Phys. Rev. Lett. 111, 101805 (2013)
\end{center}

\vspace{\fill}

{\footnotesize 
\centerline{\copyright~CERN on behalf of the \lhcb collaboration, license \href{http://creativecommons.org/licenses/by/3.0/}{CC-BY-3.0}.}}
\vspace*{2mm}

\end{titlepage}


\newpage
\setcounter{page}{2}
\mbox{~}
\newpage

\centerline{\large\bf LHCb collaboration}
\begin{flushleft}
\small
R.~Aaij$^{40}$, 
B.~Adeva$^{36}$, 
M.~Adinolfi$^{45}$, 
C.~Adrover$^{6}$, 
A.~Affolder$^{51}$, 
Z.~Ajaltouni$^{5}$, 
J.~Albrecht$^{9}$, 
F.~Alessio$^{37}$, 
M.~Alexander$^{50}$, 
S.~Ali$^{40}$
% ...

{\footnotesize \it
% ...
$ ^{5}$Clermont Universit\'{e}, Universit\'{e} Blaise Pascal, CNRS/IN2P3, LPC, Clermont-Ferrand, France\\
$ ^{6}$CPPM, Aix-Marseille Universit\'{e}, CNRS/IN2P3, Marseille, France\\
% ...
$ ^{9}$Fakult\"{a}t Physik, Technische Universit\"{a}t Dortmund, Dortmund, Germany\\
% ...
$ ^{36}$Universidad de Santiago de Compostela, Santiago de Compostela, Spain\\
$ ^{37}$European Organization for Nuclear Research (CERN), Geneva, Switzerland\\
% ...
$ ^{40}$Nikhef National Institute for Subatomic Physics, Amsterdam, The Netherlands\\
% ...
$ ^{45}$H.H. Wills Physics Laboratory, University of Bristol, Bristol, United Kingdom\\
% ...
$ ^{50}$School of Physics and Astronomy, University of Glasgow, Glasgow, United Kingdom\\
$ ^{51}$Oliver Lodge Laboratory, University of Liverpool, Liverpool, United Kingdom\\
% ...
}
\end{flushleft}

\cleardoublepage
\twocolumn


\renewcommand{\thefootnote}{\arabic{footnote}}
\setcounter{footnote}{0}

\pagestyle{plain} 
\setcounter{page}{1}
\pagenumbering{arabic}


\noindent
The rare decays \Bsmumu and \Bdmumu are highly suppressed and their branching fractions precisely predicted in the standard model (SM); any observed deviation would therefore be a clear sign of physics beyond the SM, for example a nonstandard Higgs sector.
The SM predicts branching fractions of \BRof \Bsmumu = \mbox{$(3.35 \pm
0.28) \times 10^{-9}$} and \BRof \Bdmumu = \mbox{$(1.07 \pm 0.10) \times
10^{-10}$}. These theoretical predictions are 
for decays at decay time $t=0$, and have been  
updated with respect to Refs.~\cite{Buras2012,Buras:2013uqa} using the latest 
average for the \Bs meson lifetime, $\tau_{B^0_s} = 1.516 \pm 
0.011\,\rm{ps}$~\cite{HFAG13}.
The uncertainty is dominated by the precision of lattice QCD calculations of 
the decay 
constants~\cite{Buras2012,Bazavov:2011aa,McNeile:2011ng,Na:2012kp,Laiho:2009eu}.
In the \Bs system, due to the finite width difference,  
 the comparison between the above prediction and the measured time-integrated
branching fraction requires a model-dependent
correction~\cite{deBruyn:2012wk}. The SM time-integrated prediction is 
therefore ${\cal B}(\Bsmumu) = (3.56 \pm 0.30) \times 
10^{-9}$, using the relative decay width difference $\Delta 
\Gamma_s/(2\Gamma_s) = 0.0615 \pm 0.0085$~\cite{HFAG13}.

The first search for dimuon decays of \B mesons took place 30 years 
ago~\cite{Giles:1984yg}. Since then, 
possible deviations from the SM prediction have been constrained
by various searches, with the most recent results available
in Refs.~\cite{cms2,atlas, Aaij:2012nna, PhysRevD.87.072006,
Aaltonen:2013as}. 
The first evidence for the \Bsmumu decay was reported by LHCb in 
Ref.~\cite{Aaij:2012nna}, with \BRof \Bsmumu $= (3.2^{\,+1.5}_{\,-1.2})
\times 10^{-9}$, together with the lowest limit on the \Bd decay, \BRof \Bdmumu 
$< 9.4 \times 10^{-10}$ at 95\,\% confidence level~(CL). 
The results presented in this Letter improve on and supersede our previous 
measurements~\cite{Aaij:2012nna}. They are based on data 
collected with the LHCb detector, corresponding to an integrated luminosity of 
1\invfb  of $pp$ collisions at the LHC recorded in 2011 at 
a centre-of-mass energy $\sqrt{s}=7\tev$, and 2\invfb recorded in 2012 at 
$\sqrt{s}=8\tev$. 
These data include an additional 1\invfb compared to the sample 
analysed in Ref.~\cite{Aaij:2012nna}, and have been reconstructed with 
improved algorithms and detector alignment parameters leading to slightly higher signal reconstruction efficiency and better invariant mass resolution. 
The samples from the two centre-of-mass 
energies are analysed as a combined dataset.

The analysis strategy is very similar to that employed in 
Ref.~\cite{Aaij:2012nna}, with a different multivariate operator based on a boosted decision
trees algorithm (\BDT)~\cite{Breiman,AdaBoost}.
After trigger and loose selection requirements, \mbox{\Bmumu} candidates are 
classified according to dimuon invariant mass and \BDT output. The 
distribution of 
candidates is compared with the background estimates to determine the 
signal yield and significance. The signal yield is converted into a branching 
fraction using a relative normalisation to
the channels \bdkpi and \bujpsik with $J/\psi  \to \mu^+ \mu^-$.
Inclusion of charge-conjugated processes is implied throughout this Letter. 
To avoid potential biases, candidates in the signal regions were not 
examined until the analysis procedure had been finalised. 

The \lhcb detector is a single-arm forward spectrometer covering the 
pseudorapidity range \mbox{$2<\eta<5$}, described in detail in 
Ref.~\cite{LHCbdetector}.   The simulated events used in this analysis are 
produced using the software described in Refs.~\cite{Sjostrand:2006za, 
Lange:2001uf, Allison:2006ve, *Agostinelli:2002hh, 
Golonka:2005pn, LHCb-PROC-2011-005, LHCb-PROC-2011-006}. 

Signal and normalisation candidate events are selected by a hardware trigger 
and a subsequent software trigger~\cite{LHCb-DP-2012-004}. The \Bmumu 
candidates are predominantly selected by single-muon and dimuon triggers.
Candidate \BuJpsiK decays are selected in a similar way, the only difference
being a different dimuon mass requirement in the software trigger. 
Candidate \Bhh decays (where $h^{(\prime)} = \pi, 
K$), used as control channels, are required to be 
triggered independently of the $B^0_{(s)}$ decay products.

Candidate \Bmumu decays are selected by combining two oppositely charged  
tracks with high quality muon identification~\cite{LHCb-DP-2013-001}, transverse 
momentum $p_{\rm T}$ satisfying $0.25<p_{\rm T}<40\gevc$, and momentum 
$p< 500\gevc$. The two tracks are required to form a secondary 
vertex (SV), with $\chi^2$ per degree of 
freedom less than 9, displaced from any $pp$ interaction vertex (primary 
vertex, PV) by a flight distance significance greater than 15. 
The smallest impact parameter $\chi^2$ ($\chi^2_{\rm IP}$), defined as 
the difference between the $\chi^2$ of a PV formed with and without the 
track in question, is required to be larger than 25 with respect to any PV for the muon candidates.
Only \B candidates with $p_{\rm T} > 0.5 \gevc$,  decay time less than 
$9 \times \tau_{\Bs}$~\cite{HFAG13},  
impact parameter significance $\textrm{IP}/\sigma(\textrm{IP}) < 5$ with 
respect to the PV for which the \B IP is minimal,  
and dimuon invariant mass in the 
range $[4900, 6000]\mevcc$ are selected.
The control and normalisation channels are selected with almost identical 
requirements to those applied to the signal sample. The \Bhh selection is the 
same as that of \Bmumu, except that muon identification 
criteria are not applied.  
The \BuJpsiK decay is reconstructed from a dimuon pair combined 
to form the $J/\psi \to \mu^+ \mu^-$ decay and selected in the same way as the 
\Bmumu signal samples, except for the requirements on 
the impact parameter significance and 
mass. After a requirement of $\chi^2_{\rm IP} >25$, kaon candidates are 
combined with the $J/\psi$ candidates. 
These selection criteria are completed by a requirement on the response of a multivariate operator, called MVS in  Ref.~\cite{lhcbpaper3} 
and unchanged since then, applied to candidates in both signal and normalisation channels.
After the trigger and selection requirements are applied, 55\,661 
signal dimuon candidates are found, which are used for the search.

The main discrimination between the signal and combinatorial background is 
brought by the BDT, which is optimised  using simulated samples of \Bsmm events 
for the signal and 
 \bbdim events for the background.
The \BDT combines information from the following input variables: the \B 
candidate decay time, IP and \pT; the minimum $\chi^2_{\rm 
IP}$ of the two muons with respect
to any PV; the distance of closest 
approach between the two muons; and the cosine of the angle between the
muon momentum in the dimuon rest frame and the vector perpendicular to
both the \B candidate momentum and the beam axis. Moreover two different
measures for the isolation of signal candidates are also included: the number of 
good two-track vertices a muon can make with other tracks in the event; and the 
\B candidate isolation, introduced in Ref.~\cite{cdf_iso}. 
With respect to the multivariate operator used in previous analyses~\cite{lhcbpaper3,Aaij:2012nna},
the minimum \pT of the two muons is no longer used while
four new variables are included to improve the separation power.
The first two are the absolute 
values of the differences between the pseudorapidities of the two muon 
candidates and between their azimuthal angles. The others are the angle 
of the momentum of the \B candidate in the laboratory frame, and the angle of 
the positive muon from the \B candidate in the rest frame of the \B candidate,  
both with respect to the sum of the momenta of tracks, in the rest frame of the \B 
candidate, consistent 
with originating from the decay of a $b$ hadron produced in association 
to the signal candidate.
In total, 12 variables enter into the BDT.

The variables used in the \BDT are chosen so that the dependence on 
dimuon invariant mass is linear and small to avoid biases.
The \BDT is constructed to be distributed uniformly in the range [0,1] for 
signal, and to peak strongly at zero for the background. 
The \BDT response range is divided into eight bins with boundaries 
$0.0$, $0.25$, $0.4$, $0.5$, $0.6$, $0.7$, $0.8$, $0.9$, and $1.0$. 

\begin{figure}[t]
\begin{center}
\includegraphics*[width=\columnwidth]{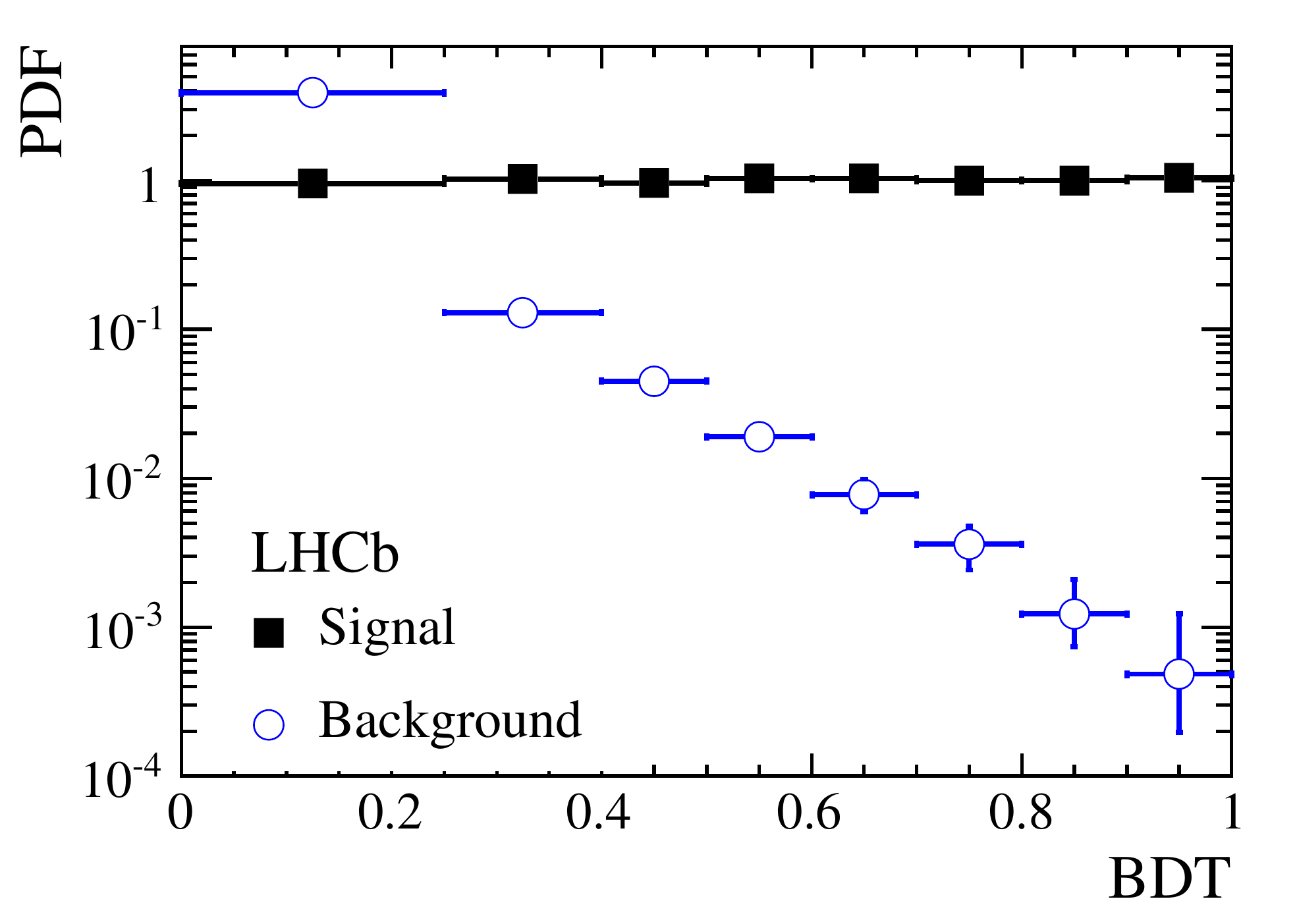}
\end{center}
\caption{\small Expected distribution of the \BDT output for the \Bsmumu signal 
(black squares), obtained from \Bhh control channels, and the  
combinatorial background (blue circles). }
\label{fig:bdt}
\end{figure}

The expected \BDT distributions for the \Bmumu signals are determined using  
\Bhh decays. The \Bhh distributions are corrected for trigger and 
muon identification distortions. An additional correction for the \Bsmumu signal 
arises from the 
difference in lifetime acceptance in \BDT bins, evaluated assuming 
the SM decay time distribution. The expected \Bsmumu \BDT distribution is shown 
in 
Fig.~\ref{fig:bdt}.

The invariant mass distribution of the signal decays 
is described by a Crystal Ball function~\cite{crystalball}. 
The peak values (\mBs and \mBd) and resolutions ($\sigma_{\Bs}$ and 
$\sigma_{\Bd}$) are obtained from \BsKK and \BdKpi, \Bdpipi decays, for the \Bs 
and \Bd mesons.
The resolutions are also determined with a power-law interpolation between the 
measured resolutions of charmonium and bottomonium resonances decaying into two 
muons. The two methods are in agreement and the combined results are 
$\sigma_{\Bs}  =  23.2 \pm 0.4 \mevcc$ and $\sigma_{\Bd}  =  22.8 \pm 0.4 
\mevcc$. The transition point of the radiative tail is obtained 
from simulated \Bsmumu events~\cite{Golonka:2005pn} smeared to reproduce the 
mass resolution measured in data.

The numbers of \Bsmumu and \Bdmumu candidates, $N_{\Bmumu}$, are converted 
into branching fractions with
\begin{eqnarray*}
\BRof \Bmumu \!\!\!\!&=&\!\!\!\!\frac{{\cal B}_{\rm norm}  \,{\rm \epsilon_{\rm 
norm}}\,f_{\rm 
norm} }{ N_{\rm norm}\,{\rm \epsilon_{sig}} \,f_{d(s)} } \times
N_{\Bmumu}  \nonumber \\
&=&\alpha^{\rm norm}_{\Bmumu} \times N_{\Bmumu},
\end{eqnarray*}
where $N_{\rm norm}$ is the number of normalisation channel decays  obtained 
from a fit to the relevant invariant mass distribution, 
and ${\cal B}_{\rm norm}$ the corresponding branching fraction. 
The fractions $f_{d(s)}$ and $f_{\rm norm}$ 
refer to the probability for a $b$ quark to fragment into the corresponding 
\B 
meson. 
The value $f_s/f_d = 0.259 \pm 0.015$, measured by LHCb in $pp$ collision data 
at $\sqrt{s}=7\tev$~\cite{Aaij:2013qqa, LHCb-CONF-2013-011}, is used and 
$f_d=f_u$ is assumed. 
The stability of $f_s/f_u$ between $\sqrt{s}=7\tev$ and $8\tev$ is verified 
by  comparing the ratios of the yields of  \BsJpsiPhi and \BuJpsiK decays. 
The effect of the measured dependence of $f_s/f_d$ on $p_{\rm 
T}$~\cite{Aaij:2013qqa} is found to be negligible.

The efficiency ${\rm \epsilon_{sig(norm)}}$ for the signal (normalisation) 
channel is the product of the reconstruction efficiency of the final state 
particles including the geometric detector acceptance, 
the selection efficiency and the trigger efficiency. 
The ratio of acceptance, reconstruction and selection efficiencies of the 
signal compared to the normalisation channel is computed 
with samples of simulated events, assuming the SM decay time distribution, corrected to take into account 
known differences between data and simulation. The tracking and 
particle identification efficiencies are measured from control channels in  
data. Residual differences between data and simulation 
are treated as sources of systematic uncertainty. 
The trigger efficiency is evaluated with data-driven 
techniques~\cite{LHCb-DP-2012-004}.

The observed numbers of \BuJpsiK and \mbox{\BdKpi} decays are
$(1.1164 \pm 0.0011)\times 10^{6}$ and $(3.76 \pm 0.06)\times 10^{4}$,
respectively.
The normalisation factors $\alpha^{\rm norm}_{\Bmumu}$ derived from the two 
 channels are consistent. Their weighted 
averages, taking correlations into account, are
$\alpha_{\Bsmumu}= (9.01 \pm 0.62) \times 10^{-11}$ and
$\alpha_{\Bdmumu}= (2.40 \pm 0.09) \times 10^{-11}$.
Assuming the \Bmumu SM branching fractions, the selected data sample is 
therefore expected to contain $40\pm4$ $B^0_{s}\to 
\mu^+\mu^-$  and  $4.5 \pm 0.4$ $B^0 \to \mu^+\mu^-$ decays in the full BDT range and with mass in $[4900, 6000]\mevcc$.

%
%
\begin{table}[b]
\tabcaption{\small Expected background yields from $b$-hadron decays, with dimuon mass 
$m_{\mu\mu}\in[4900,6000]\mevcc$ and the relative fraction with  $\rm{BDT }> 0.7$.\label{tab:backgrounds}}
\begin{center}
\begin{tabular}{l@{}rr@{$\pm$}lc}
\hline 
\Tstrut&\multicolumn{3}{c}{Yield in full}&Fraction with\\
&\multicolumn{3}{c}{BDT range}&$\rm{BDT }> 0.7$  [\%]\\
\hline 
\Tstrut\Bstrut\Bhh&&$15$&$1$&$28$\\
\Bstrut$\Bd \to \pi^- \mu^+ \nu_{\mu}$&&$115$&$6$&$15$\\
\Bstrut$B^0_s \to K^- \mu^+ \nu_{\mu}$&&$10$&$4$&$21$\\
\Bstrut$\B^{0 (+)} \to \pi^{0(+)} \mu^+ \mu^-$&&$28$&$8$&$15$\\
\Bstrut\Lbpmunu&&$70$&$30$&$11$\\ 
\hline
\end{tabular}

\end{center}
\end{table}

Invariant mass sidebands are defined as \mbox{$[4900, m_{B^0} - 60]\mevcc$} 
and \mbox{$[m_{B^0_s} + 60, 6000]\mevcc$}.
The low-mass sideband and the \Bd and \Bs signal regions contain a small amount 
of background from specific $b$-hadron decays.
A subset of this background requires the misidentification of one 
or both of the candidate muons and includes  \BdPiMuNu, 
\Bhh, $B^0_s \to K^- \mu^+ \nu_{\mu}$, and \Lbpmunu decays. 
In order to estimate the contribution from these processes, the 
$B^0 \to \pi^- \mu^+ \nu_{\mu}$ and \Bhh branching fractions are taken from 
Ref.~\cite{PDG2012}, while, in the absence of measurements, theoretical 
estimates of the \Lbpmunu~\cite{Wang:2009hra} and 
$B^0_s \to K^- \mu^+ \nu_{\mu}$~\cite{BsKmunu} branching fractions are used. 
Misidentification probabilities for the tracks in these decays are measured 
directly with 
control channels in data.
Background sources without any misidentification such as \mbox{$B^+_c 
\to J/\psi \mu^+ \nu_{\mu}$}~\cite{Abe:1998wi} and 
\mbox{$B^{0(+)} \to \pi^{0(+)} \mu^+ \mu^-$}~\cite{Bpimumu} decays are also 
considered. The expected yields of all the $b$-hadron background modes are 
estimated by normalising to the \BuJpsiK decay with the exception of \Bhh, 
for which the explicit selection yields are used, correcting for the 
trigger efficiency ratio.
No veto is imposed on photons, as the contribution of 
 $B^0_s \to \mu^+ \mu^- \gamma$ is negligible, as are contributions 
from $B^0_s \to \mu^+ \mu^- \nu_{\mu} \bar \nu_{\mu}$ 
decays~\cite{Melikhov:2004mk,Aditya:2012im}. 
The expected number of events for each of the backgrounds from $b$-hadron 
decays is shown in Table~\ref{tab:backgrounds}. 
The only one of these contributions that is relevant under the signal mass 
peaks is from \Bhh decays.

A simultaneous unbinned maximum-likelihood fit to the data is performed in 
the mass projections of the \BDT bins to determine the \Bsmumu and \Bdmumu 
branching fractions, which are free parameters.
The \Bsmumu and \Bdmumu  fractional yields in \BDT bins are constrained to 
the \BDT fractions calibrated with the \Bhh sample. 
The parameters of the Crystal Ball functions, that describe the mass shapes, 
and the normalisation factors are restricted by Gaussian constraints 
according to their expected values and uncertainties.
The backgrounds from \Bhh, $\Bd \to \pi^- \mu^+ \nu_{\mu}$, 
$B^0_s \to K^- \mu^+ \nu_{\mu}$  and $\B^{0 (+)} \to 
\pi^{0(+)} \mu^+ \mu^-$ are included as separate components in the fit. 
The fractional yields  of the $b$-hadron 
backgrounds in each \BDT bin and their overall yields 
are limited by Gaussian constraints around the expected values according to their uncertainties.
The combinatorial background in each \BDT bin is parametrised with an exponential function 
for which both the slope and the normalisation are allowed to vary freely.
The resulting \BDT distribution is compared to that expected for the signal in 
Fig.~\ref{fig:bdt}.

\begin{figure}[t]
\begin{center}
\includegraphics*[width=\columnwidth]{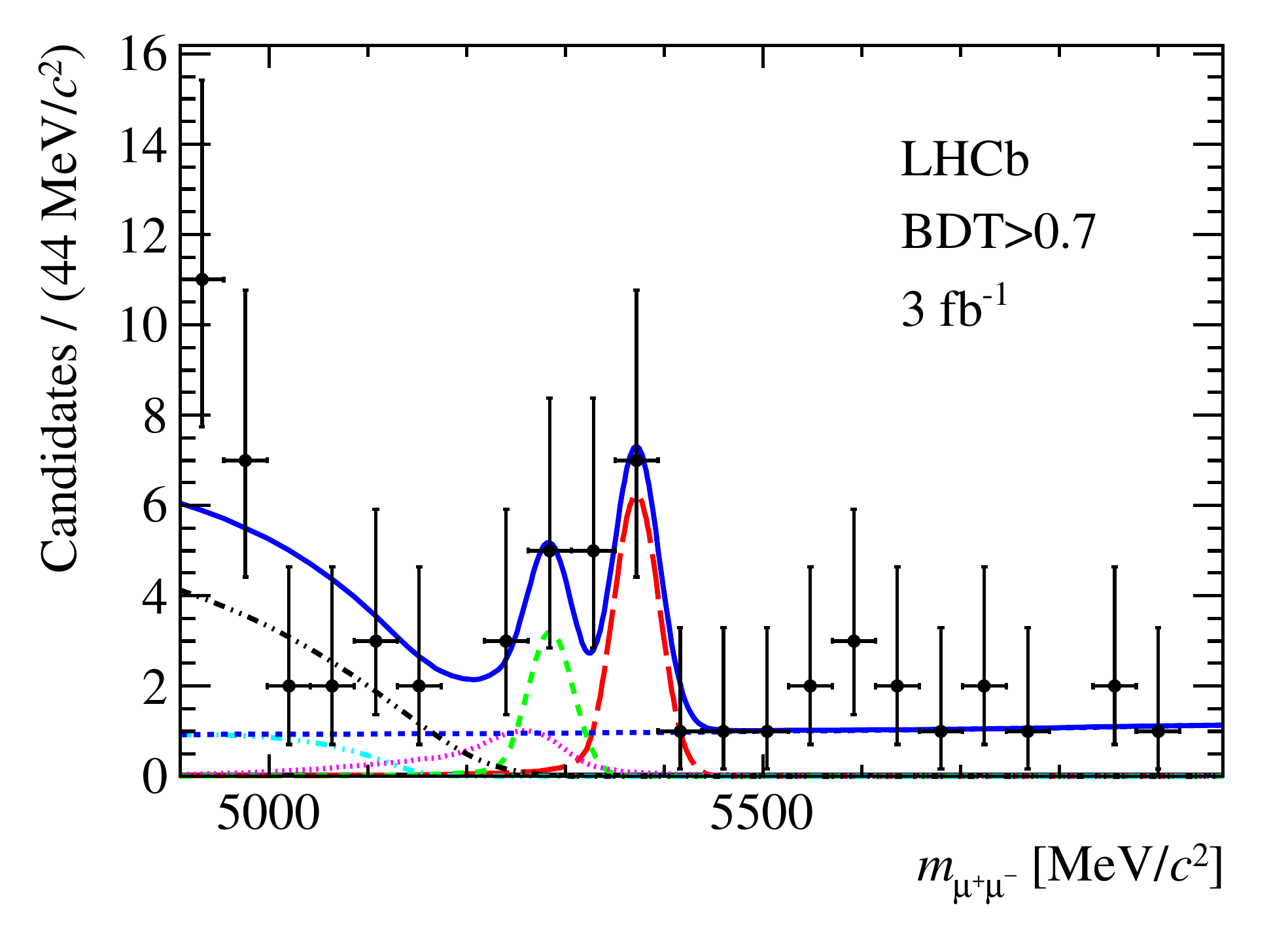}
\end{center}
\caption{\small Invariant mass distribution of the selected \Bmumu candidates (black 
dots) with ${\rm \BDT}>0.7$.
The result of the fit is overlaid (blue solid line) and the different components 
detailed: \Bsmumu  (red long dashed line), \Bdmumu (green medium dashed line), 
combinatorial background (blue medium dashed line), \Bhh (magenta dotted line), 
\bpimumu 
(light blue dot-dashed line),  \BdPiMuNu and \BsKMuNu (black dot-dashed line).}
\label{fig:mass}
\end{figure}

An excess of \Bsmumu candidates with respect to the expectation from 
the background only is seen with a significance of \Bssign standard 
deviations ($\sigma$), while 
the significance of the \Bdmumu signal is \Bdsigma. 
These significances are determined from the change in likelihood from fits with 
and without the signal component.
The median significance expected for a SM \Bsmumu signal is 
\Bsexpsigma.
  
The simultaneous unbinned maximum-likelihood fit results in 
\begin{eqnarray*}
\BRof \Bsmumu\!\!\!\!\!&=&\!\!\!\!\!\Bsbr\, ,\\
\BRof \Bdmumu\!\!\!\!\!&=&\!\!\!\!\!\Bdbr\, . 
\end{eqnarray*}
The statistical uncertainty is derived 
by repeating the fit after fixing all the fit parameters, except  the \Bsmumu 
and \Bdmumu branching fractions and the slope and normalisation of the 
combinatorial background, to their expected values.
The systematic uncertainty is obtained by subtracting in quadrature the 
statistical uncertainty 
from the total uncertainty obtained from the likelihood with all nuisance 
parameters allowed to vary
according to their uncertainties. Additional systematic uncertainties 
reflect the impact on the result of changes 
in the parametrisation of the background by including the 
\Lbpmunu component and by varying the mass shapes of backgrounds from 
$b$-hadron decays, and are added in quadrature.
The correlation between the branching fractions parameters of both decay modes is $+3.3$\,\%.
The values of the \Bmumu branching fractions obtained from 
the fit are in agreement with the SM expectations.
The invariant mass distribution of the \Bmumu candidates with
${\rm \BDT}>0.7$ is shown in Fig.~\ref{fig:mass}.

As no significant excess of \Bdmumu events is found, a modified 
frequentist approach, the \CLs method~\cite{Read_02} is used, to set an upper 
limit on the branching fraction.
The method provides \CLsb, a measure of the 
compatibility of the observed distribution with the signal plus background 
hypothesis, \CLb, a measure of the compatibility with the background-only 
hypothesis, and \mbox{$\CLs=\CLsb/\CLb$}.
A search region is defined around the \Bd invariant mass as $m_{\Bd} \pm 60\mevcc$.
For each  \BDT bin the invariant mass signal region is divided 
into nine bins with boundaries $m_{B^{0}} \pm 18, 30, 36, 48, 60\mevcc$, 
leading to a total of 72 search bins.

\begin{table}[b]
\tabcaption{\small Expected limits for the background only (bkg) and background plus SM 
signal (bkg+SM) hypotheses, and observed limits on the \Bdmumu 
branching fraction.}
\label{tab:b0limit}
\begin{center}
\begin{tabular}{lcc}
\hline 
\Tstrut      & 90\,\% \CL & 95\,\% \CL\\ 
\hline
 \Tstrut Exp. bkg        &  \Bdexpbkgn   & \Bdexpbkgnf\\ 
 \Tstrut Exp. bkg+SM     &  \Bdexpsmn  & \Bdexpsmnf\\
 \Tstrut Observed        & \Bdobslimitn & \Bdobslimitnf\\ 
\hline 
\end{tabular}
\end{center}
\end{table}

An exponential function is fitted, in each \BDT bin, to the invariant mass 
sidebands.
Even though they do not contribute to the signal search window, the $b$-hadron 
backgrounds are added as components in the fit to account for their effect on 
the combinatorial background estimate.
The uncertainty on the expected number of combinatorial background events per bin is 
determined by applying a Poissonian fluctuation to the number of events 
observed in the sidebands and by varying the exponential slopes according to 
their uncertainties.
In each bin, the expectations for \Bsmumu decays 
assuming the SM branching fraction and for \Bhh background are accounted for.
For each branching fraction hypothesis, the expected number of signal events 
is estimated from the normalisation factor. Signal events are distributed in 
bins according to the invariant mass and \BDT calibrations.
 
\begin{figure}[t]
\centering
\includegraphics[width=\columnwidth]{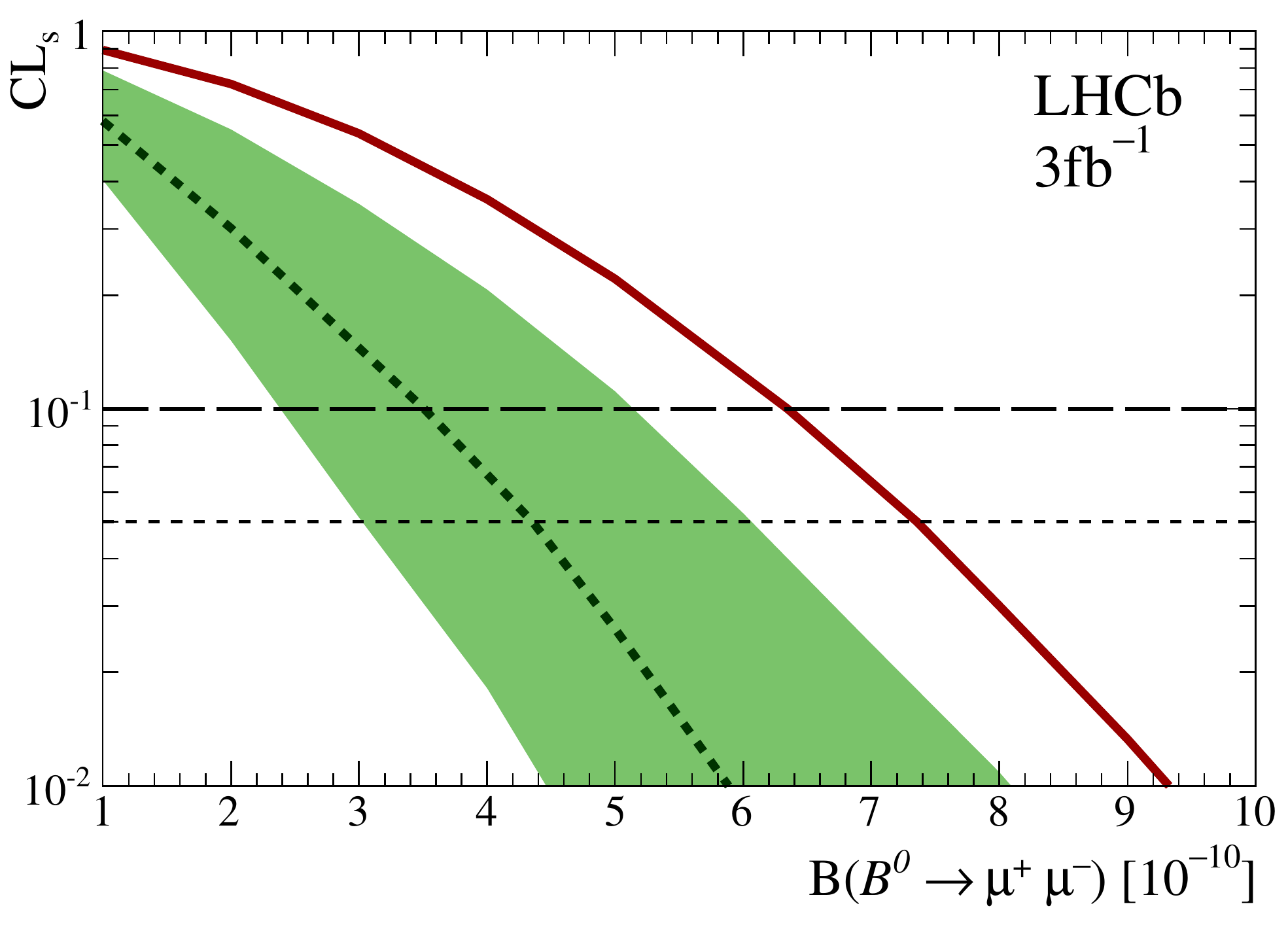}
\caption{\small\CLs as a function of the assumed \Bdmumu branching fraction. 
The dashed curve is the median of the expected \CLs distribution for 
background-only hypothesis.
The green area covers, for each branching fraction value, $34.1\,\%$ of the expected \CLs distribution on each side of its median.
The solid red curve is the observed \CLs.}
\label{fig:cls_bd}
\end{figure}

In each bin, the expected numbers of signal and background events are computed 
and compared to the number of observed candidates using \CLs.
The expected and observed upper limits for 
the \Bdmumu channel are summarised in Table~\ref{tab:b0limit} and the expected 
and observed \CLs values as functions of the branching fraction are shown in 
Fig.~\ref{fig:cls_bd}. 

In summary, a search for the rare decays \Bsmumu and \Bdmumu is performed with 
$pp$ collision data  corresponding to integrated luminosities of 1\invfb and 
2\invfb collected at $\sqrt{s} = 7\tev$ and $8\tev$, respectively.
The \Bd decay yield is not significant and an improved  upper limit of 
\mbox{$\BRof \Bdmumu < \Bdobslimitnf$} at 95\,\% CL is obtained.
The \Bsmumu signal is seen with a significance of \Bssigma. 
The time-integrated branching fraction \BRof \Bsmumu is measured to be 
\mbox{\Bsbrshort}, in agreement with the SM prediction. These measurements
 supersede and improve on our previous results, and
tighten the constraints on possible new physics contributions to these decays.

\section*{Acknowledgements}
\noindent We express our gratitude to our colleagues in the CERN
accelerator departments for the excellent performance of the LHC. We
thank the technical and administrative staff at the LHCb
institutes. We acknowledge support from CERN and from the national
agencies: CAPES, CNPq, FAPERJ and FINEP (Brazil); NSFC (China);
CNRS/IN2P3 and Region Auvergne (France); BMBF, DFG, HGF and MPG
(Germany); SFI (Ireland); INFN (Italy); FOM and NWO (The Netherlands);
SCSR (Poland); MEN/IFA (Romania); MinES, Rosatom, RFBR and NRC
``Kurchatov Institute'' (Russia); MinECo, XuntaGal and GENCAT (Spain);
SNSF and SER (Switzerland); NAS Ukraine (Ukraine); STFC (United
Kingdom); NSF (USA). We also acknowledge the support received from the
ERC under FP7. The Tier1 computing centres are supported by IN2P3
(France), KIT and BMBF (Germany), INFN (Italy), NWO and SURF (The
Netherlands), PIC (Spain), GridPP (United Kingdom). We are thankful
for the computing resources put at our disposal by Yandex LLC
(Russia), as well as to the communities behind the multiple open
source software packages that we depend on.


\ifx\mcitethebibliography\mciteundefinedmacro
\PackageError{LHCb.bst}{mciteplus.sty has not been loaded}
{This bibstyle requires the use of the mciteplus package.}\fi
\providecommand{\href}[2]{#2}

\end{document}